{  

\newcommand{\be}{\begin{equation}}
\newcommand{\bea}{\begin{eqnarray}}
\newcommand{\eea}{\end{eqnarray}}

\newcommand{\nn}{\nonumber}
\def\siml{{\ \lower-1.2pt\vbox{\hbox{\rlap{$<$}\lower6pt\vbox{\hbox{$\sim$}}}}\ }}
\newcommand{\al}{\alpha}
\def\bfnabla{\mbox{\boldmath $\nabla$}}
\def\bfsigma{\mbox{\boldmath $\sigma$}}

\newcommand{\MS}{\overline{\rm MS}}

\documentclass[a4paper,11pt]{article}

\usepackage{contribution}
\usepackage{epsfig}

\newcommand{\weblink}[2][]{%
    \ifthenelse{\equal{#1}{}}%
    {\textnormal{\url{#2}}}%
    {\textnormal{\href{#2}{#1}}}%
}

\def\beq{\begin{equation}}
\def\eeq#1{\label{#1}\end{equation}}
\def\eeqn{\end{equation}}

\def\beqa{\begin{eqnarray}}
\def\eeqa#1{\label{#1}\end{eqnarray}}
\def\eeqan{\end{eqnarray}}

\let\bar=\overbar

\def\Dslash{\not{\hbox{\kern-4pt $D$}}}
\def\dslash{\not{\hbox{\kern-2pt $\del$}}}

\def\msb{{\bar{\ssstyle M \kern -1pt S}}}

\newcommand{\contribution}[7][]{%
  \clearpage
  \thispagestyle{plain}
  \ifthenelse{\equal{#1}{}}
  {\hypersetup{pdftitle={#2}}}
  {\hypersetup{pdftitle={#1}}}
  \hypersetup{pdfauthor={{#3} {#4}}}
  {\centering\normalfont\LARGE\bfseries\sffamily #2 \par\nobreak}
  \lhead{}
  \chead{%
    \textit{\footnotesize XIV International Conference on Hadron Spectroscopy
      (\weblink[\textit{hadron2011}]{http://www.hadron2011.de}), 13-17 June 2011, Munich, Germany}%
  }
  \rhead{}
  \bigskip
  \begin{center}
    {#3} {#4}\ifthenelse{\equal{#6}{}}{}{\footnote{\weblink[#6]{mailto:#6}}}
    \ifthenelse{\equal{#7}{}}{}{#7} \\
    \textit{#5}
  \end{center}
  \bigskip
}

\renewcommand{\abstract}[1]{%
  \begin{center}
    \begin{minipage}{0.85\textwidth}
      \begin{footnotesize}
        #1
      \end{footnotesize}
    \end{minipage}
  \end{center}
  \bigskip
}

\begin{document}


\contribution[]  
{Brief review of the theory of the muonic hydrogen Lamb shift and the proton radius}  
{Antonio}{Pineda}  
{Grup de F\'\i sica Te\`orica, Universitat
Aut\`onoma de Barcelona, E-08193 Bellaterra, Barcelona, Spain}  
{}  
{}  

%

\abstract{%
Recently the muonic hydrogen lamb shift has been measured with unprecedented accuracy, allowing for a precise determination of the proton radius. This determination is 5 sigma away from the previous CODATA value obtained from (mainly) the hydrogen lamb shift and the electron-proton scattering. Within an effective field theory formalism, I will define the proton radius and briefly review some aspects of the theoretical prediction for the muonic hydrogen lamb shift, studying both the pure QED-like computation and the hadronic effects. 
  }
%


\section{Introduction}

The recent measurement of the muonic hydrogen Lamb shift, $E(2P_{3/2}(F=2))-E(2S_{1/2}(F=1))$,
$$
E_{exp}=206.2949(32) {\rm meV}
$$
and the associated determination of the electromagnetic proton radius\cite{Pohl:2010zza}:
\be
\label{rpexp}
r_p= 0.84184(67) {\rm fm}
\,.
\end{equation}
has led to a lot of controversy. The reason is that this number is 5 sigma away from the CODATA value, $r_p= 0.8768(69)$ fm \cite{Mohr:2008fa}. If 
instead one uses this value in the theoretical expression of the muonic Hydrogen Lamb shift one gets the following discrepancy:
\begin{equation}
E_{exp}-E_{th}=0.311  \; {\rm meV}
\label{difference}
\end{equation}
between theory and experiment. Two main options are clearly at hand: either the theoretical determination is not correct (or not as precise as claimed), or 
previous determinations of the proton radius were incorrect (or not as precise as claimed). Here we would like to study the theoretical expression of the muonic Hydrogen Lamb shift within an effective field theory perspective. We do it partially, and only focus on some few aspects, as a full analysis would require much more space. In particular spin effects will not be considered. We believe that the use of effective field theories helps in organizing the computation by providing with power counting rules to asses the importance of the different contributions. This will be even more important once higher order effects are included. For the present discussion a ${\cal O}(m_r\alpha^5)$ precision is enough to visualize the discrepancy. Higher order effects are way smaller than the discrepancy found in Eq. (\ref{difference}). Moreover ${\cal O}(m_r\alpha^5)$ is the only thing completely known at present\footnote{This is also the precision presently reached in heavy quarkonium spectrum computations. This made provide with some cross checks between both systems.}.

The dynamics of the muonic hydrogen is characterized by several scales:\\
$m_p \sim \Lambda_{\chi}$,\\
$m_{\mu} \sim m_{\pi} \sim m_r=\frac{m_{\mu}m_p}{m_p+m_{\mu}}$,\\
$m_r\al \sim m_e$.\\
By considering ratios between them the main expansion parameters are obtained:\\
$\displaystyle{\frac{m_{\pi}}{m_p} \sim \frac{m_{\mu}}{m_p} \sim \frac{1}{9}}$,\\
$\displaystyle{\frac{m_r\al}{m_r}\sim \frac{m_r\al^2}{m_r\al}\sim \al \sim\frac{1}{137}}$.\\

We use the effective field theory Potential Non-Relativistic QED (pNRQED) \cite{Pineda:1997bj}. Specially relevant for us is Ref. 
\cite{Pineda:2004mx}, which contains much more detailed information on the application of pNRQED to the muonic hydrogen, and we refer to it for details 
(see also \cite{Pineda:1998kn,Pineda:2002as,Nevado:2007dd}). pNRQED profits from the hierarchy  $m_{\mu} \gg m_{\mu}\al \gg m_{\mu}\al^2$ and the Lagrangian reads
\bea
\label{lpnrqed}
&&L_{pNRQED} =
\int d^3{\bf r} d^3{\bf R} dt S^{\dagger}({\bf r}, {\bf R}, t)
                \Biggl\{
i\partial_0 - { {\bf p}^2 \over2 m_r} 
\\
&&
\nonumber
- V ({\bf r}, {\bf p}, {\bfsigma}_1,{\bfsigma}_2) + e {\bf r} \cdot {\bf E} ({\bf R},t)
\Biggr\}
S ({\bf r}, {\bf R}, t)- \int d^3{\bf r} {1\over 4} F_{\mu \nu} F^{\mu \nu}
\,,
\eea
where $S$ is the field representing the muonic hydrogen, ${\bf R}$ the center of mass coordinate and ${\bf r}$ the relative distance.

$V$ stands for the potential and admits an expansion in powers of $1/m_{\mu}$:
\be
V ({\bf r}, {\bf p}, {\bfsigma}_1,{\bfsigma}_2)
=
V^{(0)}(r)+{V^{(1)}(r) \over m_{\mu}}+{V^{(2)}(r) \over m_{\mu}^2}+\cdots\,.
\end{equation}
The potential is obtained through matching to the underlying theory. Since pNRQED describes degrees of freedom with $E \sim m_{\mu}\al^2$, 
any other degree of freedom with larger energy is integrated out. This implies treating the proton and muon in a non-relativistic fashion and integrating out 
pions. This is the step of going from Heavy Baryon Effective Theory (HBET) to NRQCD. By integrating out the scale $m_{\mu}\al$, pNRQED is obtained 
and the potentials appear. Schematically the path followed is the following: 
$$
HBET (m_\pi/m_{\mu}) \rightarrow NRQED (m_{\mu}\al) \rightarrow pNRQED\,.
$$

\section{Pure QED contributions}

We first focus on the pure QED contributions. We mostly follow Pachucki's work \cite{pachucki1,pachucki2,pachucki4}, as it mainly follows an strict order by order in $\al$ computation, trying to accommodate their results in our formalism. See however \cite{Pohl:2010zza} (or \cite{Borie2011,Jentschura:2010ej}) for more complete list of references.

The static potential can be written in the following way in momentum space
\begin{equation}
{\tilde V}^{(0)} \equiv  - 4\pi Z_{\mu}Z_p\alpha_{V}(k){1 \over {\bf k}^2},  
\end{equation}
\be
\alpha_{eff}(k)=\al{1 \over 1+\Pi(-{\bf k}^2)}
\,,
\end{equation}
where $\Pi(k^2)$ is the vacuum polarization due to electrons and can be computed order by order in $\al$:
$$
\Pi(k^2)=\al\Pi^{(1)}(k^2)+\al^2\Pi^{(2)}(k^2)+\al^3\Pi^{(3)}(k^2)+...
$$ 
\be 
\alpha_{V}(k)=
\alpha_{eff}(k)+\sum_{n,m=0 \atop n+m=even>0} Z_{\mu}^nZ_p^m\al_{eff}^{(n,m)}(k) 
=\alpha_{eff}(k)+\delta \al(k)\,, \quad \quad \delta \al(k)={\cal O}(\al^4)
.
\end{equation}
\begin{center}
\begin{figure}[!htb]
\vspace*{-25ex}
\hspace*{-2ex}\epsfxsize=20truecm \epsfbox{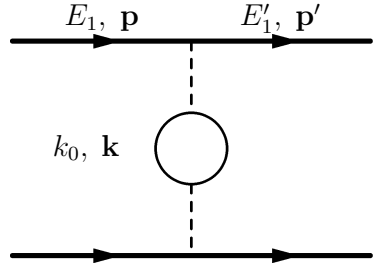}
\vspace*{-117ex}
\caption{\it Leading correction to the Coulomb potential due to the electron vacuum
  polarization. ${\bf k}={\bf p}-{\bf p}^{\prime}$ and 
  $k_0=E_1-E_1^{\prime}$.}
\label{figvacuum}
\end{figure}
\end{center}

The leading order contribution to the lamb shift comes from the one-loop vacuum polarization correction to the static potential (see Fig. \ref{figvacuum})
$$
E_{LO}=\langle n|\delta V|n\rangle=205.0074 \; {\rm mev}={\cal O}(m_r\al^3)\,.
$$

\begin{center}
\begin{figure}[!htb]
\vspace*{-4ex}
\hspace*{16ex}\epsfxsize=10truecm \epsfbox{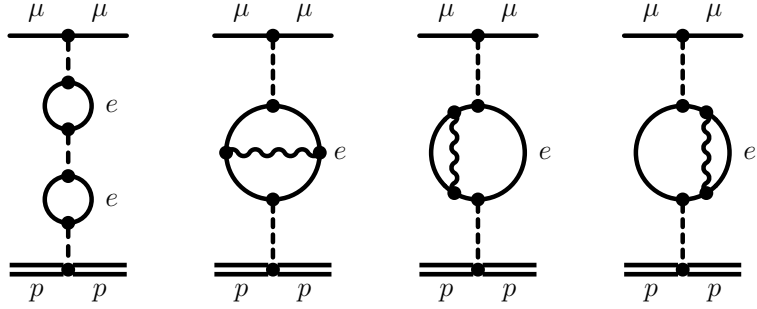}
\vspace*{-1ex}
\caption{\it The static potential at two loops.}
\label{twoloop}
\end{figure}
\end{center}

The ${\cal O}(m_r\al^4)$ contribution to the lamb shift comes from the two-loop static potential and from 
the iteration of the one-loop potential in quantum mechanics perturbation theory. The latter yields 
$\Delta E = 0.151 \; {\rm meV}$. The former is purely due to vacuum polarization corrections (see Fig. \ref{twoloop}) and yields
$\Delta E = 1.5079 \; {\rm meV}$.

The three-loop static potential contribution due to the vacuum polarization (and the associated iterations from perturbation theory 
were computed in Ref. \cite{Kinoshita:1998jf} (see also \cite{Ivanov:2009aa} for an small correction).
The result was quite small $\Delta E = 0.0076 \; {\rm meV}={\cal O}(m_r\al^5)$.

All previous contributions were due to the vacuum polarization. The only contribution from the static potential 
that  is not due to the vacuum polarization at ${\cal O}(m_r\al^5)$ comes from $\delta \alpha$. It is a  light-by-light (Wichmann-Kroll and Delbr\"uck) contribution 
and very small \cite{Karshenboim:2010cq}
$$
\Delta E \simeq -0.0009 \; {\rm meV}
\,.
$$
It should be mentioned that the limit $m_e \rightarrow 0$ of the static potential is known at three loops from QCD \cite{Anzai:2009tm,Smirnov:2009fh}, which could be used as a check. It is also reasonable to think that the result with finite $m_e$ could also be obtained from these results (albeit numerically) with a finite amount of work.

There are no corrections due to the $1/m_{\mu}$ potential at ${\cal O}(m_r\al^5)$. From the $1/m^2_{\mu}$ potential 
(see \cite{Pineda:2004mx} for its expression in pNRQED) there are the tree level relativistic corrections, which give 
$\Delta E = 0.0575 \; {\rm meV}={\cal O}(m\al^4)$. The incorporation of the one-loop vacuum polarization to the relativistic 
$1/m^2_{\mu}$ tree-level potential gives the following 
result  $\Delta E = 0.0169 \; {\rm meV}={\cal O}(m\al^5)$  \cite{pachucki4}.

\begin{center}
\begin{figure}[!htb]
\vspace*{-1ex}
\hspace*{30ex}\epsfxsize=4truecm \epsfbox{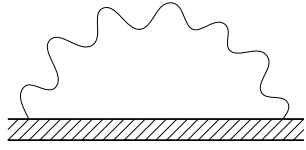}
\caption{\it self-energy correction to the muonic hydrogen energy due to the interaction with ultrasoft photons.}
\label{figUS}
\end{figure}
\end{center}

In order to complete the pure QED ${\cal O}(m\al^5)$ corrections one has to include the interaction with the ultrasoft photons (see Fig. \ref{figUS}). 
They yield the result (taken from \cite{pachucki1})
$$
\Delta E=-0.6677\; {\rm meV}={\cal O}(m\al^5)
\,.
$$
The $\frac{m_{\mu}}{m_p}$ ultrasoft effects contribute
$$
\Delta E=-0.045\; {\rm meV}={\cal O}(m\al^5\frac{m_{\mu}}{m_p})
\,.
$$
In pNRQED these results would not come from the interaction with the ultrasoft photons only, as it would be factorization scale dependent, they also include effects due to the NRQED matching coefficients encoded in the $1/m^2_{\mu}$ potentials. The procedure is 
pretty much the same the one used for positronium in Ref. \cite{Pineda:1998kn}. The details for muonic hydrogen will be worked out elsewhere.

At ${\cal O}(m\al^5\frac{m^2_{\mu}}{m^2_p})$ one starts to have overlap with hadronic effects, which we discuss in the next section.

\section{Hadronic Contributions}
In the previous section we have considered the proton to be point-like. We now incorporate the finite-size effects 
due to the hadronic structure of the proton. These effects are encoded in the coefficient multiplying the delta potential 
(note that the combination of NRQED matching coefficients that appears in the potential is always the same).
\be
\delta V_{had}^{(2)}(r) \equiv  \frac{1}{m_p^2}D_d^{had}\delta^3({\bf r})
 \rightarrow  \Delta E = \frac{1}{m_p^2}D_d^{had}\frac{1}{\pi}(\frac{m_r\al}{n})^3\delta_{l0}
\end{equation}
where
\be
D_d^{had}=-c^{had}_3-16\pi\al d^{had}_2+ \frac{\pi\al}{2}c^{had}_D
\,.
\end{equation}
We define
$c_3=c^{point-like}_3+c^{had}_3$, $d_2=d_2^{point-like}+d^{had}_2$
$c_D=c_D^{point-like}+c_D^{had.}$ ,so that
 $c^{had}_3$, $d^{had}_2$, $c^{had}_D$ are the left-over of the 
matching coefficients of NRQED Lagrangian 
\be
\delta {\cal L}
=\cdots \frac{d_2}{m_p^2}F_{\mu\nu}D^2F^{\mu\nu}+\cdots
-e\frac{c_D}{m_p^2}N_p^{\dagger}\bfnabla \cdot {\bf E}N_p
+\cdots+
\frac{c_3}{m_p^2}N^{\dagger}_pN_p\mu^{\dagger}\mu
\end{equation}
after subtraction of the point-like contributions.
We do in this way because traditionally the point-like contributions are already included in the "pure" QED corrections 
described in the previous section\footnote{Note though, that for an strict effective theory point of view, at scales of the order of $m_p$, 
it is not a good approximation to consider the proton point like. Therefore, in a way, we are introducing an "spurious" contribution in the 
hadronic matching coefficients.}. I more extended discussion can be found in Refs. \cite{Pineda:2004mx,Pineda:2002as}.

$d^{had}_2$ encodes the hadronic vacuum polarization effect. Its contribution to the Lamb shift is tiny, $\Delta E=0.011\;{\rm meV}$, 
and not much subject to uncertainty as it can be determined with enough precision from dispersion relations. 

\begin{figure}[h]
\makebox[3.8cm]{\phantom b}
\epsfxsize=7.5truecm \epsfbox{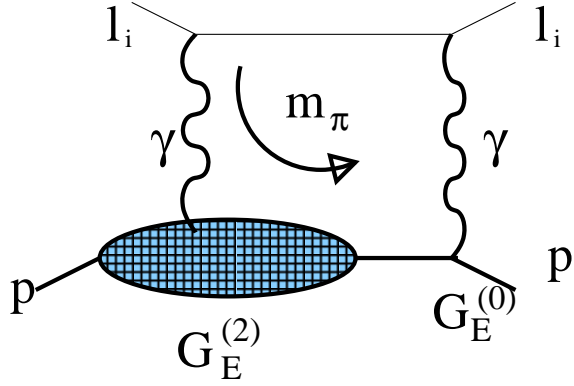}
\caption {{\it Symbolic representation (plus permutations) 
of the Zemach $\langle r^3 \rangle$ correction, Eq. (\ref{c3Zemach}).}}
\label{figzemach}
\end{figure}

\begin{figure}[h]
	\centering
		\includegraphics[width=7.5cm]{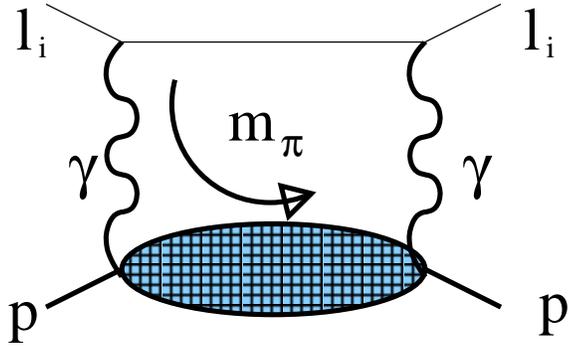}
\caption{\label{figc3}{\it Symbolic representation of Eq. (\ref{c3pol}).}
}
\end{figure}

More subject to discussion are the hadronic corrections associated to $c_3^{had}$. They are usually split into two terms (see the discussion in 
Refs. \cite{Pineda:2004mx,Pineda:2002as}): $c^{had}_3=c^{had}_{3,Zemach}+c^{had}_{3,pol}$. We symbolically draw them in Figs. \ref{figzemach} and \ref{figc3}, and discuss 
them in the next subsections. A common feature of both of them is that they are power-like chiral enhanced: $\sim \frac{m_{\mu}}{m_{\pi}}$. This is very important, as 
it allows chiral perturbation theory to predict the leading order term without introducing any extra parameter. The resulting correction to 
the Lamb shift is of ${\cal O}(m_{\mu}\al^5\times\frac{m_{\mu}^2}{\Lambda^2_{\chi}}\times \frac{m_{\mu}}{m_{\pi}})$. 


\subsection{Zemach correction, $\langle r^3 \rangle$}
It is the one analogous to the Zemach correction defined in 
the hyperfine splitting \cite{Zemach:1956zz}. It is also common to rewrite it in terms of a coefficient $\langle r_p^3 \rangle$
\be
\label{c3Zemach}
c^{had}_{3,Zemach}= \frac{\pi}{3}\alpha^2 m_p^2m_{\mu}\langle r_p^3 \rangle
\,,
\qquad
\frac{\langle r_p^3 \rangle}{\rm fm^3}
=
\frac{96}{\pi}\int {d^{D-1}k}
{1 \over {\bf k}^6}G_E^{(0)}G_E^{(2)}
\,,
\end{equation}
where $G_E^{(n)}$ is the electric Sachs form factor to order $n$ in the chiral counting ($G_E^{(0)}=1$). We also use dimensional 
regularization ($D=4+2\epsilon$). This gets rid of power-like divergences which are then automatically set to zero (no need for counterterms as 
one would with cutoff regularization). The final result is finite and it is possible to obtain an analytic expression for the leading term in the chiral and large $N_c$ expansion (by including the $\Delta$ particle contribution). It reads \cite{Pineda:2004mx}
\bea
c^{had}_{3,Zemach}&=& 
{2}(\pi\alpha)^2 
\left({m_p \over 4\pi F_0 }\right)^2
{m_{\mu} \over m_\pi}
\left\{
{3 \over 4}g_A^2+{1 \over 8}
\right.
\\
&&
\left.
+{2 \over \pi}g_{\pi N\Delta}^2{m_\pi \over \Delta}
\sum_{r=0}^{\infty}C_r\left({m_\pi \over \Delta}\right)^{2r}
+g_{\pi N\Delta}^2\sum_{r=1}^{\infty}
H_r\left({m_\pi \over \Delta}\right)^{2r}
\right\}
\,,
\nn
\eea
where ($\Delta=M_{\Delta}-M_p \sim 300$ MeV)
\be
C_r={(-1)^r\Gamma(-3/2) \over \Gamma(r+1)\Gamma(-3/2-r)}
\left\{
B_{6+2r}-{2(r+2) \over 3+2r}B_{4+2r} 
\right\}
\,,
\qquad r \geq 0
\,,
\end{equation}
\be
B_n \equiv \int_0^\infty \,dt
{t^{2-n} \over \sqrt{1-t^2}}\ln{\left[{1\over t}+\sqrt{{1 \over t^2}-1}\right]}
\qquad
H_n\equiv {n!(2n-1)!!\Gamma[-3/2] \over 2(2n)!!\Gamma[1/2+n]}  
\,.
\end{equation}
This expression produces the following number for $\langle r_p^3 \rangle$ and the associated energy shift:
\be
\frac{\langle r_p^3 \rangle|_{\chi PT}}{\rm fm^3}
=
1.9 \; ({\rm Pineda})\rightarrow \Delta E =
0.010
\frac{\langle r_p^3 \rangle}{\rm fm^3}=0.019 \; {\rm meV}
\end{equation}
This number can be compared with some recent determinations of $\langle r_p^3 \rangle$ using dispersion relations 
\cite{Friar:2005jz,Distler:2010zq,Bernauer:2010wm,Arrington:2007ux} 
\begin{eqnarray}
\nonumber
 \frac{\langle r_p^3 \rangle|_{"exp"}}{\rm fm^3} = \,\left\{
\begin{array}{ll}
 \displaystyle{2.71(13)\;{\rm Friar-Sick}}
 &
\\
 \displaystyle{2.85(8)\;{\rm  Bernauer-Arrington}} 
 &
\end{array}
\right\} \rightarrow \Delta E=0.027-0.029
\end{eqnarray}
In principle the difference between these two determinations comes from different fit functions and data, 
which may give a first estimate of the associated uncertainty of the dispersion relation analysis. We find quite reassuring that the 
difference with the chiral computation is around 40 \%, which could be easily accommodated with higher order corrections. Much more difficult to 
accommodate would be the value advocated in Ref. \cite{DeRujula:2010dp}, $\langle r_p^3 \rangle \sim 36.5$, from a direct fit to the muonic hydrogen Lamb shift using the CODATA value for the proton value. This would require that higher order corrections in the chiral computation to be a factor 15 larger than the leading order result. We believe this is at odds with chiral symmetry, even more so taking into account that one of the motivations of such proposal was the lack of experimental data at low momentum, but it is precisely in this region where chiral perturbation theory should work better. 

\subsection{Polarizability correction}

The determination of the polarizability correction from experiment is on more shaky grounds than for the Zemach correction, producing the larger uncertainty 
in the theoretical expression for the Lamb shift. The reason is that dispersion relations do not fix the result completely. The final number used in \cite{Pohl:2010zza} was taken from the average in Ref. \cite{Borie:2004fv} $\Delta E=0.015 \pm 0.004$ using the results from, \cite{pachucki2} $\Delta E=0.012\pm 0.002$ meV, \cite{Rosenfelder:1999px} $ \Delta E=0.017 \pm 0.004\; {\rm meV}$, and \cite{Faustov} $ \Delta E=0.016\; {\rm meV}$. For a recent discussion see 
Ref. \cite{Carlson:2011zd}.

Here again chiral computations may turn out to be crucial to asses the size of this correction. The reason, as before, is that the polarizability correction is 
power-like chiral enhanced. Therefore, chiral perturbation theory can predict the leading order term with no new parameter. This is the attitude followed in Ref. \cite{Nevado:2007dd}, where a chiral computation using dispersion relations yielded
\bea
&&
\nn
c_{3,pol}^{had}=- e^4 m_pm_{\mu}\int {d^4k_E \over (2\pi)^4}{1 \over k_E^4}{1 \over
k_E^4+4m_{\mu}^2k_{0,E}^2 }
\left\{
(3k_{0,E}^2+{\bf k}^2)S_1(ik_{0,E},-k_E^2)-{\bf k}^2S_2(ik_{0,E},-k_E^2)
\right\}
\eea
where
\begin{equation}  \nn
 T^{\mu\nu} = i\!\int\! d^4x\, e^{iq\cdot x}
  \langle {p,s}| {T J^\mu(x)J^\nu(0)} |{p,s}\rangle
\,,
\end{equation}
which has the following structure ($\rho=q\cdot p/m$):
\bea \nn
 T^{\mu\nu} &=&
  \left( -g^{\mu\nu} + \frac{q^\mu q^\nu}{q^2}\right) S_1(\rho,q^2) 
   + \frac1{m_p^2} \left( p^\mu - \frac{m_p\rho}{q^2} q^\mu \right)
    \left( p^\nu - \frac{m_p\rho}{q^2} q^\nu \right) S_2(\rho,q^2) \nn \\
  && - \frac i{m_p}\, \epsilon^{\mu\nu\rho\sigma} q_\rho s_\sigma A_1(\rho,q^2)
     - \frac i{m_p^3}\, \epsilon^{\mu\nu\rho\sigma} q_\rho
   \bigl( (m_p\rho) s_\sigma - (q\cdot s) p_\sigma \bigr) A_2(\rho,q^2)
\eea

\begin{figure}[h]
	\centering
		\includegraphics[width=9.5cm]{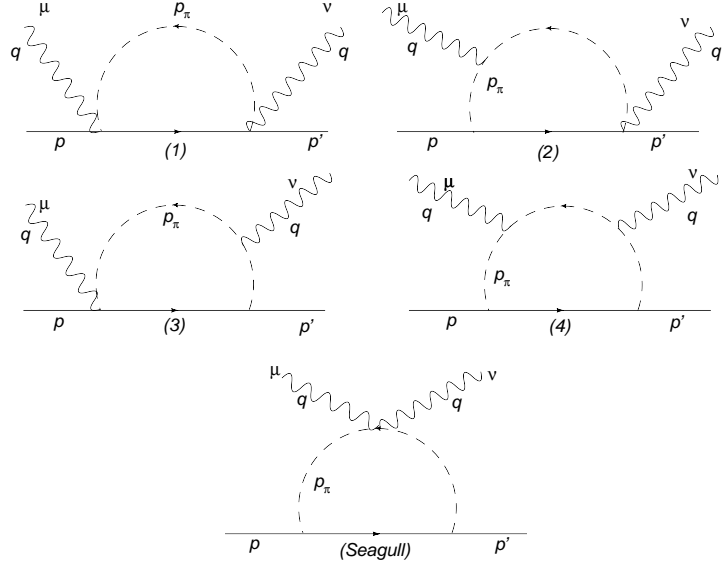}
\caption{\label{figTij}Diagrams contributing to $T^{ij}$. Crossed diagrams are not 
explicitly shown but calculated. 
}
\end{figure}

After introducing the chiral expressions for the structure factors from the diagrams in Fig. \ref{figTij}, one obtains
\bea
\label{c3pol}
&&
c_{3,pol}^{had}=- e^4 m_p^2\frac{m_{\mu}}{m_{\pi}}\left(\frac{g_A}{f_{\pi}}\right)^2
\int {d^{D-1}k_E \over (2\pi)^{D-1}}{1 \over (1+{\bf k}^2)^4}
\int_0^{\infty}\frac{dw}{\pi}w^{D-5}{1 \over
w^2+4\frac{m^2_{\mu}}{m_{\pi}^2}\frac{1}{(1+{\bf k}^2)^2}}
\\
\nn
&&
\times
\left\{
(2+(1+{\bf k}^2)^2)A_E(w^2,{\bf k}^2)+(1+{\bf k}^2)^2{\bf k}^2w^2B_E(w^2,{\bf k}^2)
\right\}
\eea
where (for $D=4$)
\bea
&&
A_E=-\frac{1}{4\pi}
\left[-\frac{3}{2}+\sqrt{1+w^2}
+\int_0^1dx\frac{1-x}{\sqrt{1+x^2w^2+x(1-x)w^2{\bf k}^2}}
\right]
\,,
\eea
\bea
\nn
&&
B_E=\frac{1}{8\pi}
\left[
\int_0^1dx\frac{1-2x}{\sqrt{1+x^2w^2+x(1-x)w^2{\bf k}^2}}
-
\frac{1}{2}\int_0^1dx\frac{(1-x)(1-2x)^2}{(1+x^2w^2+x(1-x)w^2{\bf k}^2)^{\frac{3}{2}}}
\right]
\,.
\eea
This gives the number
\be
\label{chiralpol}
\Delta E|_{\chi PT}(pions)=0.018 \; {\rm meV}
\,.
\end{equation}
We consider a more through chiral study of this object, in particular including the $\Delta$ particle, compulsory. The introduction of the $\Delta$ particle 
produced a large effect in the case of the Zemach correction, something similar may happen here.  Whereas we can (and should, see Ref. \cite{Hill:2011wy} for a recent discussion) further analyze the error associated to the polarizability correction, we would like to emphasize that in order to explain Eq. (\ref{difference}), the corrections to the leading order chiral computation should be a factor 15 larger than the number obtained in Eq. (\ref{chiralpol}).

\subsection{Definition of the proton radius}
From the effective theory point of view, the proton radius corresponds to an specific combination of the Wilson coefficients of the 
effective theory. Let us see how this relation appears. One first considers the following matrix element
\be
\langle {p^\prime,s}|J^\mu|{p,s}\rangle
=
\bar u(p^\prime) \left[ F_1(q^2) \gamma^\mu +
i F_2(q^2){\sigma^{\mu \nu} q_\nu\over 2 m_p} \right]u(p)
\label{current}
\,.
\end{equation}
We are interested in the low energy limit of the form factors
\be
F_i(q^2)=F_i+{q^2 \over m_p^2}F_i^{\prime}+...\,
\end{equation}
or more precisely of the Sachs form factors
\be
G_E(q^2)=F_1(q^2)+{q^2 \over 4m_p^2}F_2(q^2), \qquad G_M(q^2)=F_1(q^2)+F_2(q^2). 
\end{equation}
The proton radius is usually defined as the derivative of the Electric Sachs form factor at zero momentum:
\be
\displaystyle{r_p^2(\nu)=6\frac{d}{dq^2}G_{E,p}(q^2)|_{q^2=0}}
=\frac{3}{4}\frac{1}{m_p^2}
\left(c_D^{(p)}(\nu)-1\right)
\end{equation}
\be
c_D=1+2F_2+8F_1^{\prime}=1+8m_p^2\left.{d G_{p,E}(q^2) \over d\,q^2}\right|_{q^2=0}
\,,
\end{equation}
This set of equations allows to visualize the relation between the proton radius and the matching coefficients of the effective theory \cite{Pineda:2004mx}. 
They also highlight the problem of defining the proton radius through the derivative of the Electric Sachs form factor at zero momentum, as this object is infrared divergent. 
Then, what is the definition of the proton radius used to obtain Eq. (\ref{rpexp})? It corresponds to subtract the point-like effect of $c_D$:
$$
r_p^2=\frac{3}{4}\frac{1}{m_p^2}
\left(c_D(\nu)-c_{D,point-like}(\nu)\right)
$$
where, in the $\MS$ scheme,
$$
c_{D,point-like}=1+\frac{\al}{\pi}\left(\frac{4}{3}\ln\frac{m_p^2}{\nu^2}\right)
\,.
$$

\section{Conclusions}

We have briefly reviewed the theoretical determination of the spin-independent corrections to the Lamb shift that contribute at ${\cal O}(m_r\al^5)$. 
We believe that it is important to have a model independent and efficient approach to the problem. Effective Field Theories are suitable for this task. 
The use of effective theories highlights that the proton radius is a matching coefficient of the effective theory and, in general, an scheme/scale dependent object. 
In principle, this is not a problem as far as one knows the definition one is using.

The pure QED computation appears to be solid, not to say extremely reliable. Yet it could be interesting to reanalyze some parts of the computation from an effective theory perspective. We also remark that the correction coming from the three-loop static potential has only been computed by one group. On the other hand, the analogous QCD static potential has been obtained by two groups. We believe that with a reasonable amount of effort, such computations may yield cross checks of the QED computation for muonic hydrogen.

For the hadronic corrections there are precise determinations of the effects associated to  the 
Zemach correction, $\langle r^3 \rangle$, using dispersion relations. Chiral perturbation theory provides with a highly non-trivial double check of the magnitude of this correction. The reason is that the chiral computation is power-like chiral enhanced. It actually linearly  diverges in the chiral limit. 
Therefore, the leading order computation in chiral perturbation theory is a pure prediction, with no free parameter. This rules out much larger values of $\langle r^3 \rangle$ than those obtained from experiment, as such values would be in tension with the chiral perturbation theory prediction.

The polarizability correction is the major source of uncertainty. The reason is that dispersion relations alone are not able to fully determine this quantity, suffering from some ambiguity in the parameterization. Therefore, the chiral perturbation theory result may turn out to be crucial here to determine the size of this correction. Again the chiral computation is power-like chiral enhanced and linearly diverges in the chiral limit. 
Thus, the leading order computation in chiral perturbation theory is a pure prediction, with no free parameter. At present there is room for improvement over the result obtained in Ref. \cite{Nevado:2007dd} using chiral perturbation theory with dispersion relations. In particular, it does not include the contribution due to the $\Delta$ particle, which, in the case of the Zemach term, turned out to be quite important. It will then be very important to compute it to really asses the size of this correction. Obviously any eventual determination from lattice of this quantity would be most welcome. 

\medskip

{\bf Acknowledgments}. This work was partially supported by the spanish 
grants FPA2007-60275 and FPA2010-16963, and by the catalan grant SGR2009-00894.


%


\end{document}